\documentclass{appolb}
\usepackage{authblk} 
\usepackage{graphicx}
\usepackage{enumitem}
\usepackage{amsmath} 
\usepackage{url}
\usepackage{mcite}
\usepackage{lineno}
\Preprinttrue
\begin{document}

\date{22 July 2025}

\setlist{noitemsep, topsep=0pt}

\title{Prospects for Higgs Boson Research at the LHC\thanks{Presented at the 10$^{\rm th}$ International Conference on Precision Physics and Fundamental Physical Constants
(FFK-2025),
Warsaw, Poland, 26-30 May, 2025.
}%
}
\author{Andr\'e Sopczak\\
on behalf of the ATLAS and CMS collaborations
\address{Institute of Experimental and Applied Physics \\ 
Czech Technical University in Prague\\
Prague, Czech Republic}
}
\maketitle
\begin{abstract}
The search for Higgs bosons in the Standard Model (SM) of particle physics and Beyond the Standard Model (BSM) started intensively at the Large Electron-Positron (LEP) collider, which operated from 1989 to 2000, and later at the Tevatron from 2001 to 2011. In 2012, with the discovery of a Higgs boson at the Large Hadron Collider (LHC) at CERN, a new era began. This allowed for precision measurements of the Higgs boson properties which, so far, are all consistent with the SM expectations. Many searches for predicted BSM Higgs bosons advanced the field of experimental Higgs boson physics. The LHC already operated in three running periods: Run-1 from 2010 to 2012, Run-2 from 2015 to 2018, and currently Run-3 from 2022 to 2026. The High-Luminosity LHC (HL-LHC) operation is foreseen from 2030. The prospects of experimental Higgs boson research for the next decade are reviewed.
\end{abstract}

\tableofcontents
  
\section{Introduction}
\begin{figure}[htbp]
\vspace{-2mm}
\includegraphics[width=0.655\textwidth]{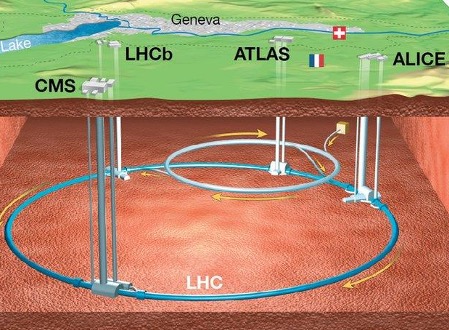}
\includegraphics[width=0.335\textwidth]{figures/hl-lhc2}
\caption{\label{fig:lhc}
Left: Illustration of the LHC tunnel. Right: Photo of current construction work in preparation for the HL-LHC.
\vspace{-6mm}
}
\end{figure}

After the Higgs boson was proposed in 1964 by Higgs~\cite{Higgs1964}, Englert and Brout~\cite{BroutEnglert1964}, and Guralnik, Hagen and Kibble~\cite{GHK1964}, an era for Higgs boson searches started at the Large Electron-Positron (LEP) collider  which operated from 1989 to 2000 in two phases, LEP-1 at 91\,GeV centre-of-mass energy and then later at LEP-2 with centre-of-mass energies up to 209~GeV. Searches at LEP-1 were summarized in several articles, for example~\cite{Sopczak:547938}. The LEP era reached sensitivity for the Standard Model (SM) Higgs boson up to 114.4~GeV at 95\% Confidence Level (CL), when the results from the four LEP experiments ALEPH, DELPHI, L3 and OPAL were combined~\cite{Barate:610122}.
The LEP collaborations performed several searches for Beyond the Standard Model (BSM) Higgs bosons~\cite{Sopczak:2006hep}. A new era of Higgs boson searches started at the Tevatron collider at Fermilab with the CDF and D0 experiments. The combination of their datasets led to evidence of the Higgs boson in the $H\rightarrow bb$ channel just before the Large Hadron Collider (LHC) era at CERN started~\cite{Sopczak:2824511}. In 2012, the SM Higgs boson was discovered at the LHC by the ATLAS and CMS experiments~\cite{Aad:1471031,CMS2012}. 
Since then, at the LHC, several precision measurements were performed for the SM Higgs boson as briefly summarized in~\cite{Sopczak:2708121} and detailed in~\cite{Aad:2814946,Tumasyan:2814513}.
Several searches were conducted for the BSM Higgs bosons as reviewed, for example, in~\cite{Sopczak2024}. In this review, the prospects of Higgs boson physics at the High-Luminosity LHC (HL-LHC), which will start operating from 2030, are reviewed. The LHC tunnel system and a view of the current preparation for the HL-LHC are shown in Fig.~\ref{fig:lhc}.

Major open questions will be addressed in future Higgs boson analyses~\cite{Salam2022}:
\begin{itemize}
    \item Why is the electroweak interaction so much stronger than gravity?
    \item Why is there more matter than antimatter in the Universe?
    \item What is dark matter?
    \item What is the origin of the vast range of quark and lepton masses?
    \item What is the origin of the early Universe inflation?
\end{itemize}

This review is based on studies for the European Strategy for Particle Physics (ESPP)~\cite{EuropeanStrategy2020}, which is the cornerstone of Europe’s decision-making process for the long-term future of the field. It was mandated by the CERN Council, and it is formed through a broad consultation of the particle physics community.  Opinions of physicists from around the world were actively solicited, and it was developed in close coordination with similar processes in the US and Asia to ensure coordination between regions and optimal use of resources globally. The first update was adopted in 2013, the second was approved in 2020, and the third update is expected to conclude in January 2026.
The prospects of Higgs boson research at the LHC are based on 
\begin{itemize}
\item ATLAS and CMS data analyses on full LHC Run-2 dataset, 2015-2018, with about 140 fb$^{-1}$ per experiment, and 
\item some of the anticipated improvements for Run-3, 2022-2026, with larger than $ 170$\,fb$^{-1}$ per experiment.
\end{itemize}

\begin{figure}[htbp]
\vspace{-5mm}
\begin{center}
\includegraphics[width=0.7\textwidth]{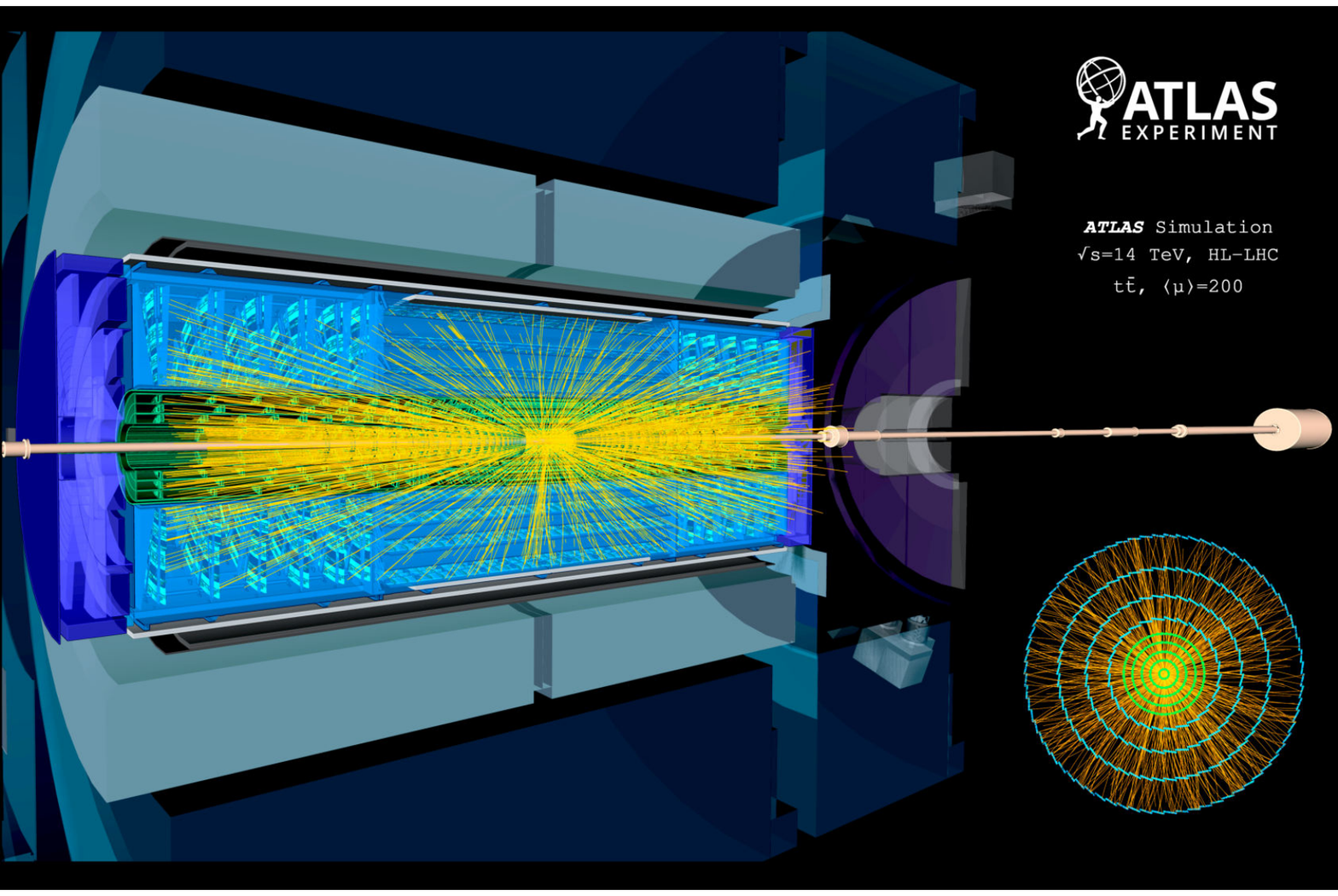}
\end{center}
\vspace{-6mm}
\caption{\label{fig:tracker}
ATLAS simulation of the HL-LHC operation. From~\protect\cite{ATLASPHOTO2023}.}
\end{figure}

The estimated prospects take into account the ATLAS and CMS detector upgrades for the HL-LHC. The detectors will have to cope with up to 200 proton-proton interactions per bunch crossing. This is a challenge for the detector operation and event reconstruction, as illustrated in a simulated top-quark pair-production ATLAS event, showing the tracker in the HL-LHC operation (Fig.~\ref{fig:tracker}~\cite{ATLASPHOTO2023}).

Considerations for the extrapolated precisions are based on two different per-experiment integrated luminosity ($\cal L$) scenarios at centre-of-mass energy at $\sqrt{s} =14$\,TeV:
\begin{itemize}
    \item nominal 3\,ab$^{-1}$ and 
    \item intermediate 2\,ab$^{-1}$. 
\end{itemize}
This assumes the data are good for physics analysis, i.e. after taking into account detector recording inefficiencies and data quality requirements, typically accounting for a 10\% decrease in integrated luminosity.
Whenever possible, combined ATLAS and CMS results are reported. In cases where results are available from only one collaboration, they are assumed to apply to both. 
Insights gained from a decade of detector operations are taken into account.
Extensive studies on the impact of the HL-LHC detector upgrades were performed.
It is assumed that ATLAS and CMS experiments should achieve comparable sensitivity.
The first scenario is referred to as S2 for backward compatibility with the past 2020 European strategy and the naming convention assumes reduced systematic uncertainties in most cases, with theoretical systematic uncertainties halved.
The additional scenario (S3) evaluates the impact of recent improvements in specific object reconstruction relative to S2, focusing on analyses that have already demonstrated and documented advancements compared with the latest publications~\cite{ATL-PHYS-PUB-2025-018}.

\section{Measurement of Higgs boson coupling to fermions and vector bosons
}
\begin{figure}[htbp]
\begin{center}
\includegraphics[width=0.44\textwidth]{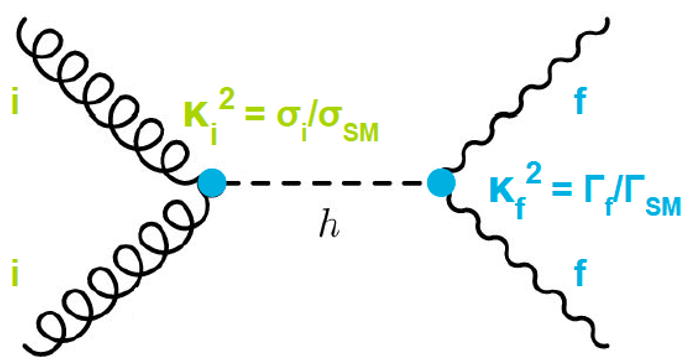}
\includegraphics[width=0.28\textwidth]{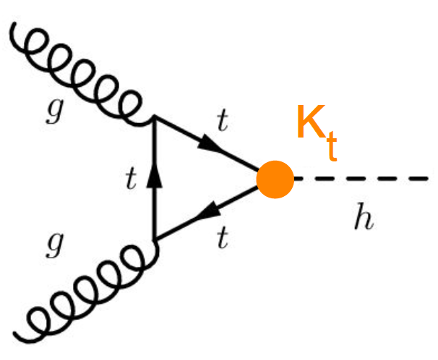}
\includegraphics[width=0.26\textwidth]{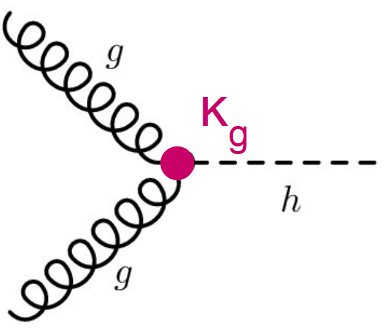}
\end{center}
\vspace{-5mm}
\caption{\label{fig:graph}
Coupling definitions for Higgs boson production and decay.}
\vspace{-7mm}
\end{figure}

\begin{figure}[htbp]
\begin{center}
\includegraphics[width=0.49\textwidth]{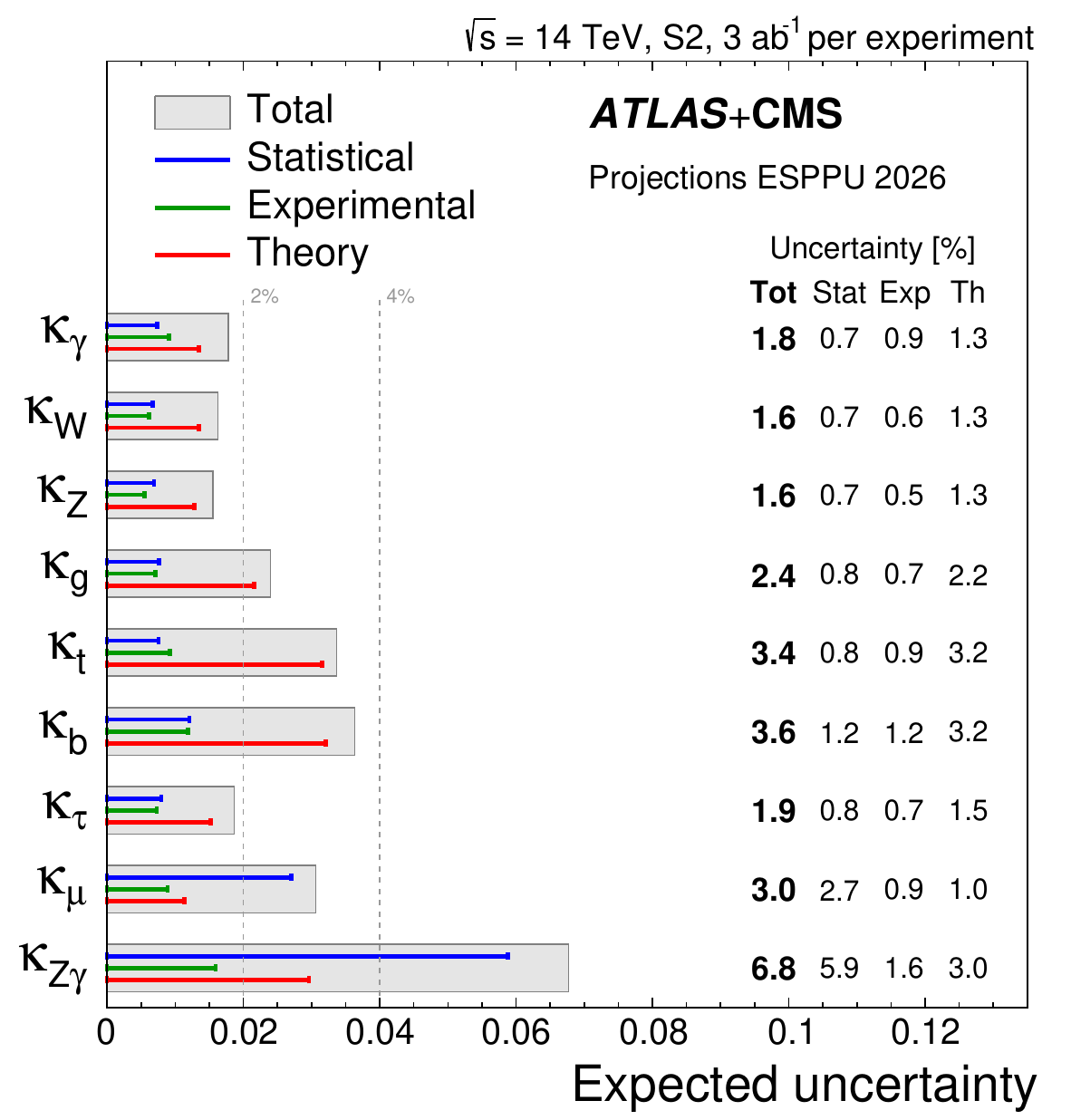}
\includegraphics[width=0.50\textwidth]{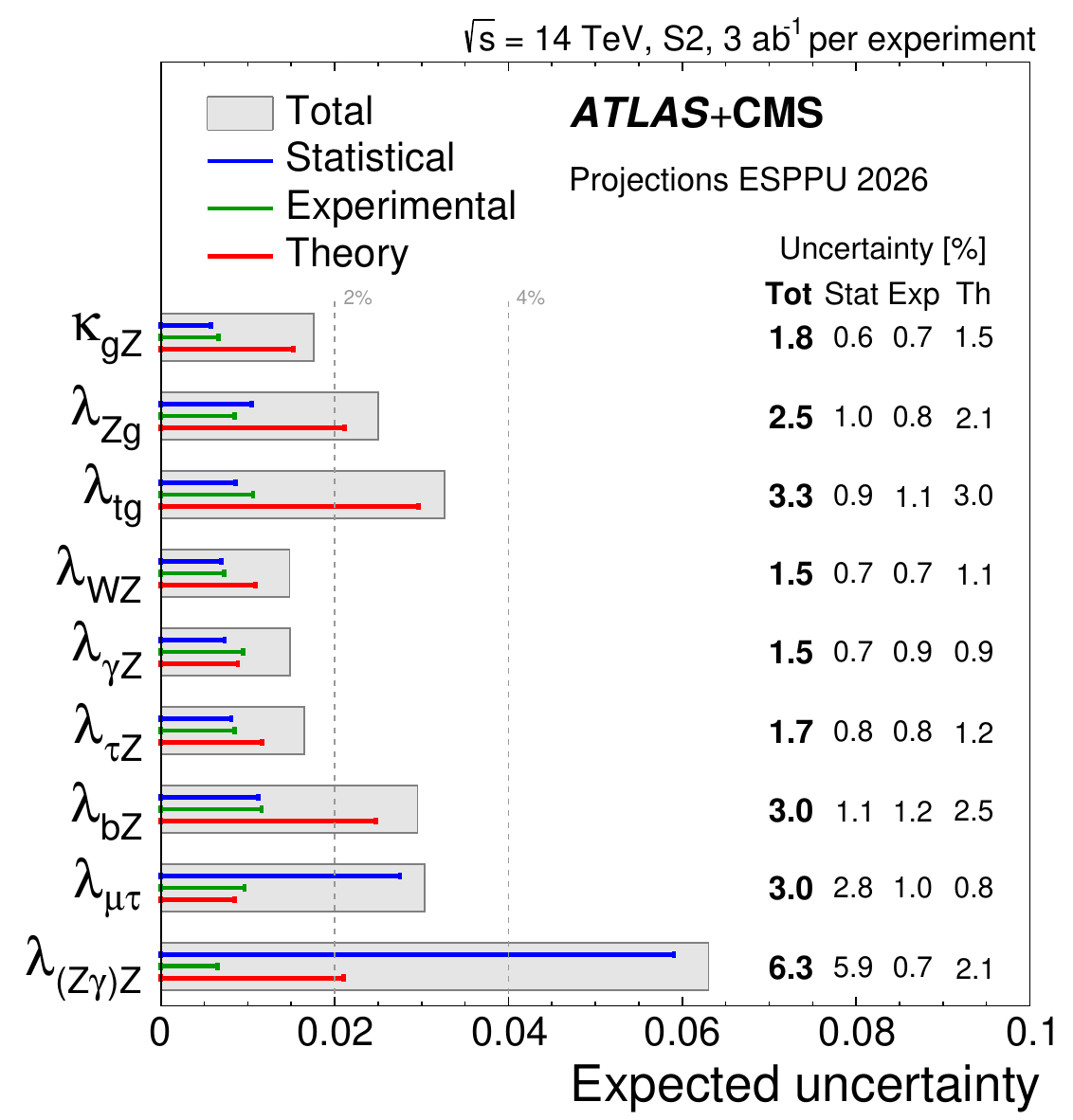}
\end{center}
\vspace{-7mm}
\caption{\label{fig:couplings}
Expected uncertainties for $\kappa$ and $\lambda$ measurements.
From~\protect\cite{ATL-PHYS-PUB-2025-018}.
\vspace{-6mm}
}
\end{figure}

For the measurement of BSM Higgs boson couplings to fermions and vector bosons (including loop-induced couplings to photons and gluons), the kappa framework is used and lambda is defined as $\lambda_{xy}=\kappa_x/\kappa_y$.
The cross-section $\sigma_H$ and branching ratio $B(H\rightarrow ff)$ at Leading-Order (LO) are proportional to the square of the effective Higgs boson coupling to the SM particles. 
In order to search for BSM effects regarding Higgs boson production and decay, $\kappa$ factors are defined $\sigma_i B_f =\sigma_i^{\rm SM} B_f^{\rm SM}\kappa_i^2\kappa_f^2$. Figure~\ref{fig:graph} illustrates the couplings.
The projected uncertainties of the coupling signal strength modifiers, $\kappa$, and their ratios $\lambda$, are shown in Fig.~\ref{fig:couplings}~\cite{ATL-PHYS-PUB-2025-018} for expected combined ATLAS and CMS results. The contributions from statistics, experimental, and theory uncertainties, as well as the total uncertainties are given.
It is remarkable that theory uncertainties are often dominant.

\clearpage

\vspace{-7mm}
\section{Observation of Higgs boson rare processes
}
The future measurement of rare Higgs boson decay modes could give first evidence of BSM physics.
The uncertainties on signal strengths $\mu$ defined as the measured value with respect to the SM expectation for the rare decays $H\rightarrow Z\gamma$ and $H \rightarrow \mu\mu$ are determined (Table~\ref{tab:rare}~\cite{ATL-PHYS-PUB-2025-018}). High precision $\mu$ measurements can be achieved both in the 2\,ab$^{-1}$ and 3\,ab$^{-1}$ scenarios.

\begin{table}[htbp]
    \centering
    \begin{tabular}{l|c|c|c}
    & & \multicolumn{2}{c}{$\delta \mu$ [\%]}\\
    $\cal L$  & &$H\rightarrow Z\gamma$ & $H \rightarrow \mu\mu$\\ \hline
    & ATLAS & 21 & 13 \\ 
    2\,ab$^{-1}$& CMS & 22 & 8.4 \\ 
    & ATLAS+CMS & 15 & 7.1 \\ \hline
    & ATLAS & 17 & 11 \\ 
   3\,ab$^{-1}$ & CMS & 19 & 7.0 \\ 
    & ATLAS+CMS & 14 & 5.9 \\ \hline    
    \end{tabular}
    \caption{Uncertainties on signal strengths $\mu$ defined as the measured value with respect to the SM expectation for the rare decays $H\rightarrow Z\gamma$ and $H \rightarrow \mu\mu$. From~\protect\cite{ATL-PHYS-PUB-2025-018}.}
    \label{tab:rare}
\end{table}

\section{Observation of the SM di-Higgs boson production}
\begin{figure}[htbp]
\begin{center}
\includegraphics[width=0.59\textwidth]{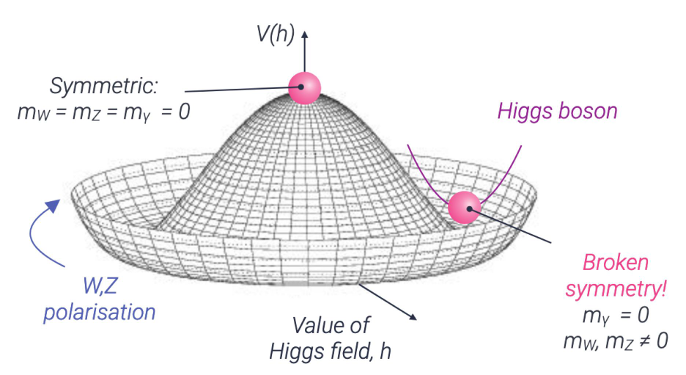}
\includegraphics[width=0.39\textwidth]{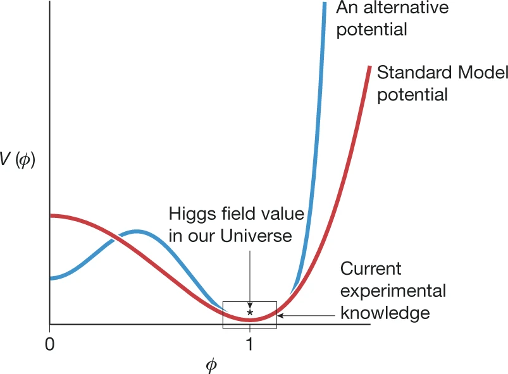}
\includegraphics[width=0.27\textwidth]{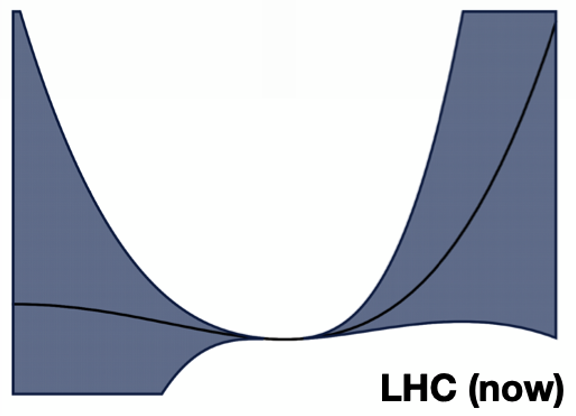}
\includegraphics[width=0.3\textwidth]{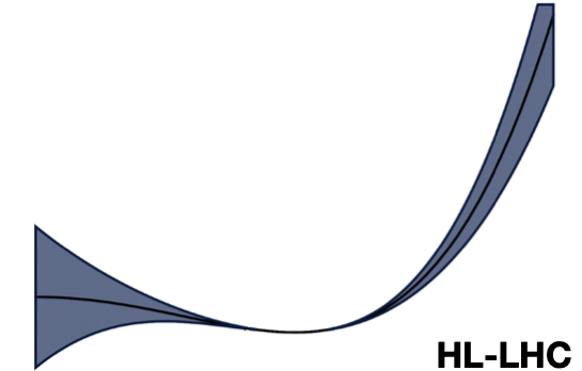}
\end{center}
\caption{\label{fig:potential}
Upper: Higgs boson potential, and SM and alternative potentials. 
Lower: Illustration of current and future uncertainties (shaded regions) on the Higgs boson potential measurements.}
\vspace{-5mm}
\end{figure}

A central question in Higgs boson physics is related to the Higgs boson potential. 
In particular, a key question is this: is the current minimum the true minimum — and thus the Universe is stable — or is there a deeper one elsewhere, allowing for decay?
Figure~\ref{fig:potential} illustrates the Higgs boson potential for the SM potential and an alternative potential representing an unstable vacuum. The figure also illustrates the uncertainty on the potential measurement from the current LHC results and future expectations at the HL-LHC.

\subsection{Measurement of the Higgs boson trilinear self-coupling $\lambda_3$}
\begin{figure}[htbp]
\begin{center}
\includegraphics[width=0.49\textwidth]{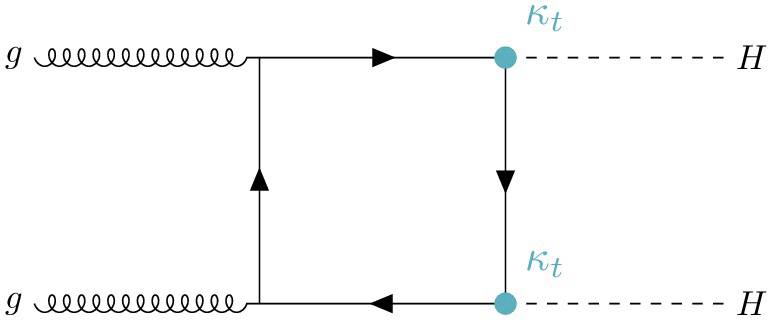}
\includegraphics[width=0.49\textwidth]{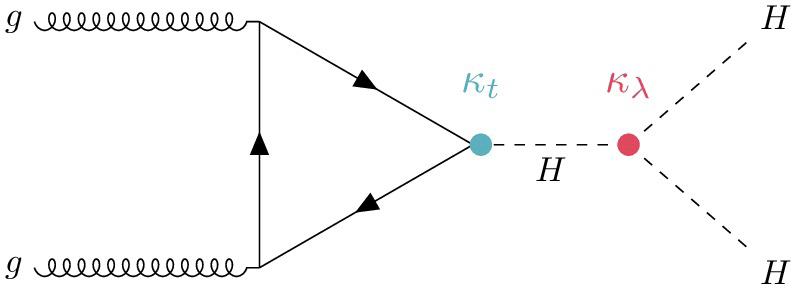}
\end{center}
\vspace{-5mm}
\caption{\label{fig:dihiggs}
Di-Higgs boson Feynman diagrams.
}
\end{figure}

\begin{figure}[htbp]
\begin{center}
\includegraphics[width=0.56\textwidth]{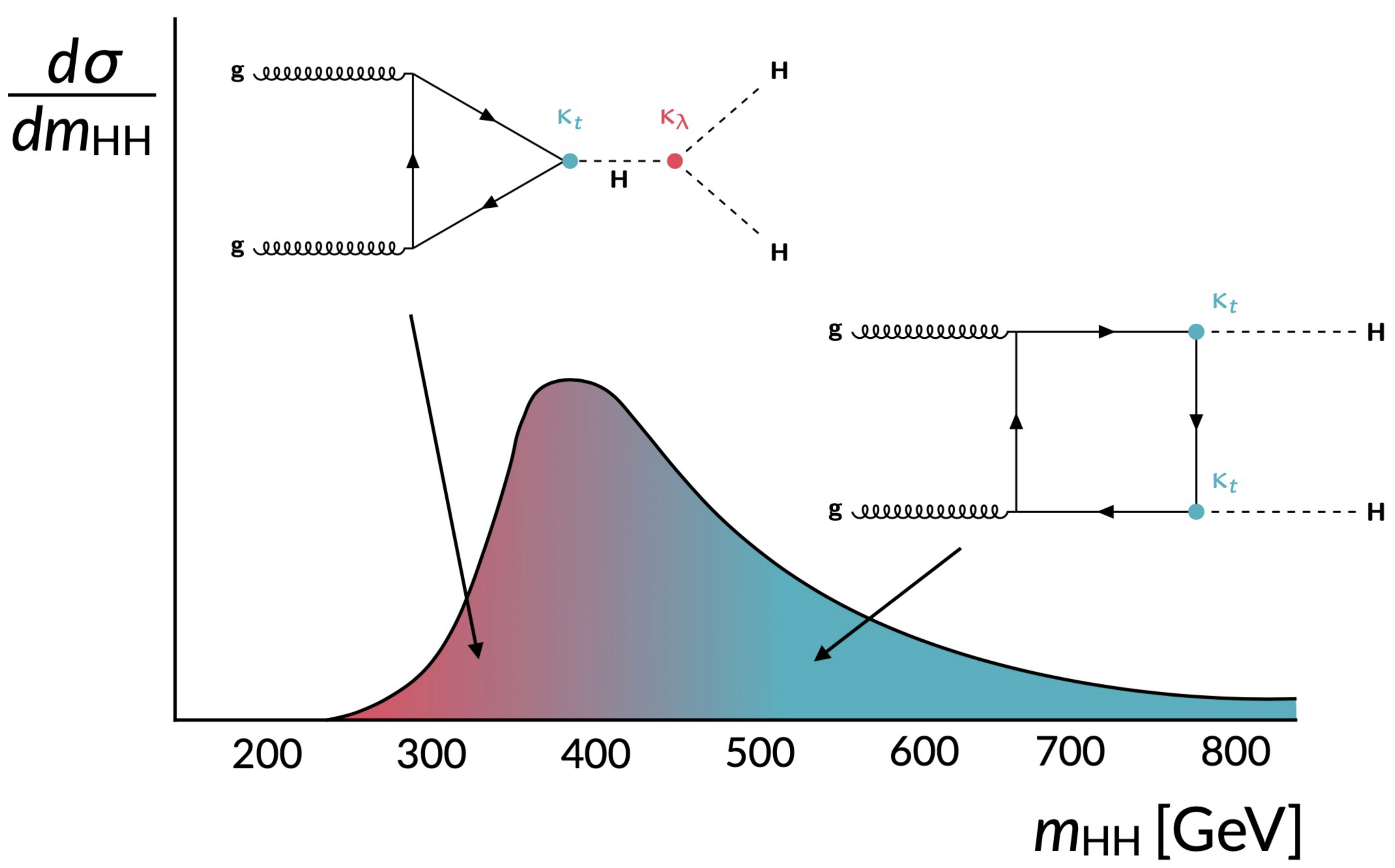}
\includegraphics[width=0.43\textwidth]{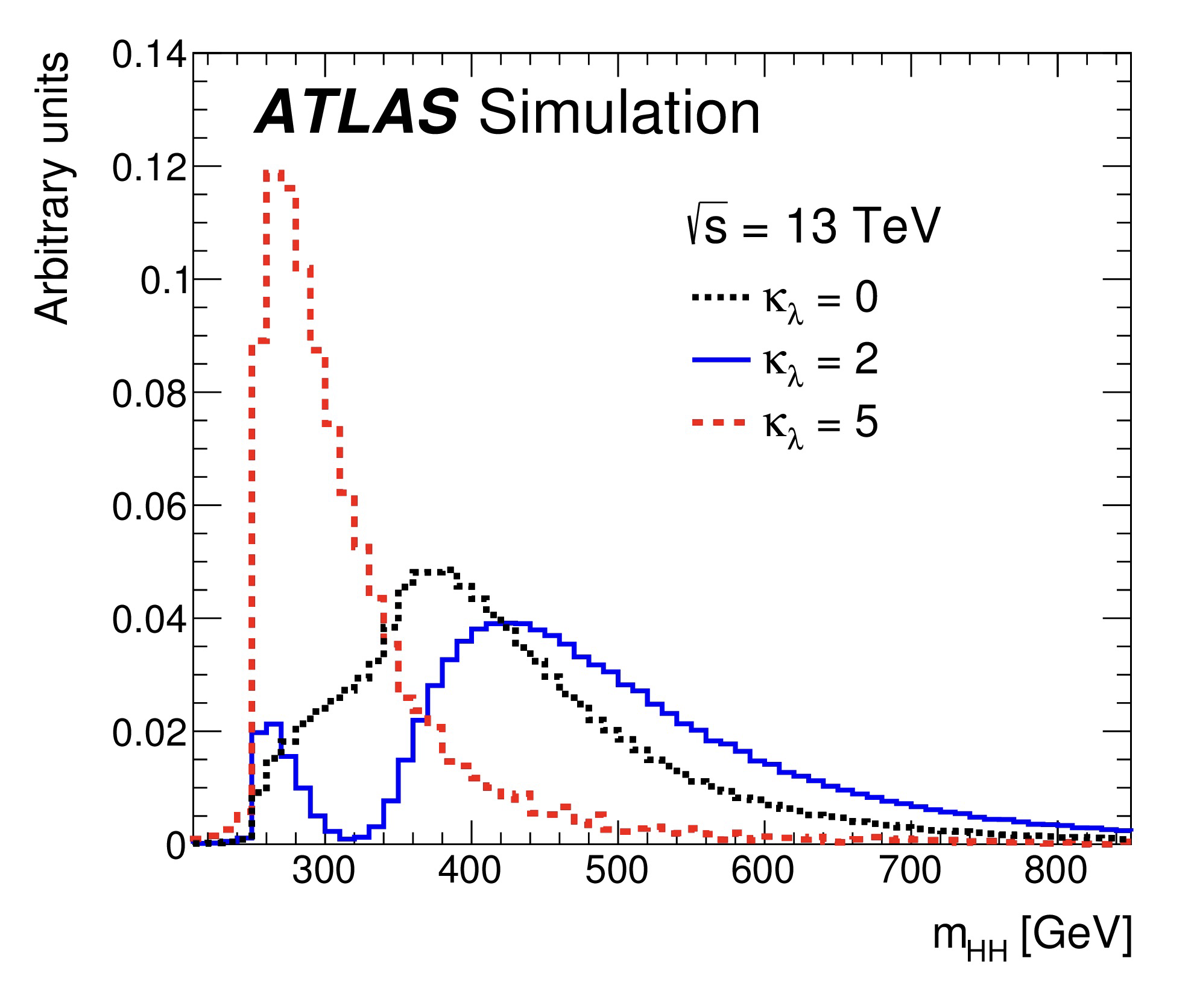}
\includegraphics[width=0.55\textwidth]{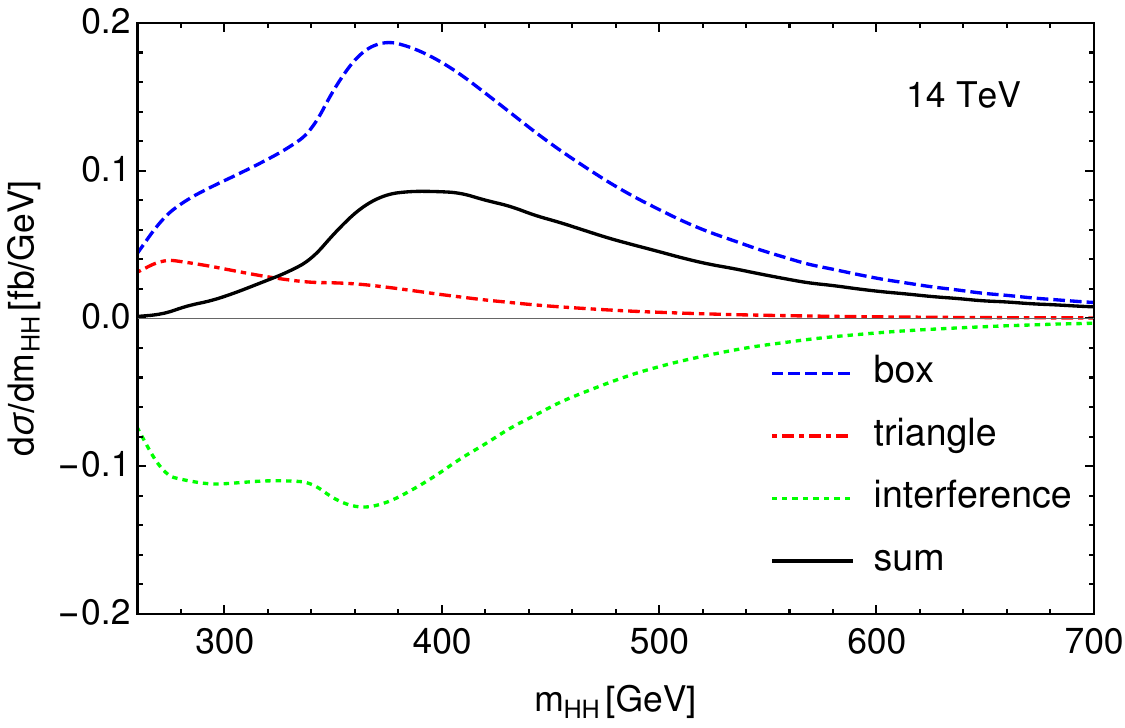}
\end{center}
\vspace{-7mm}
\caption{\label{fig:boxtriangle}
Di-Higgs expected differential cross-section as a function of the di-Higgs boson (HH) mass.
From~\protect\cite{Collaboration:2759320}.
}
\end{figure}

The Higgs potential can be investigated by measuring the Higgs boson self-coupling $\lambda$, using di-Higgs boson production. In the SM, $\lambda_3= m_H^2/(2v^2)$, where $v$ is the Vacuum Expectation Value (VEV), 246 GeV. The $\lambda$  value is an essential parameter in the Higgs boson potential
\begin{equation}
V(h) = \frac{1}{2} m_H^2h^2 +  \lambda_3 v h^3 + \frac{1}{4}\lambda_4 h^4.
\end{equation}
A di-Higgs boson signature can arise from the Feynman diagrams shown in Fig.~\ref{fig:dihiggs}.
For 
$\kappa_\lambda=\lambda_3(HHH)/\lambda_3(HHH)_{\rm SM}=0$, only the box diagram contributes.
Maximum destructive interference is obtained at $\kappa_\lambda \simeq 2$.
For $\kappa_\lambda > 5$, the triangle diagram dominates.
This relation is shown in Fig.~\ref{fig:boxtriangle}~\cite{Collaboration:2759320}.
The triangle diagram contributes for low HH masses, while for high HH masses, the box diagram dominates. In arbitrary units, an ATLAS simulation shows the expected spectrum of contributions for different $\kappa_\lambda$ values as a function of the HH mass. 
Derivations from the SM value, $\kappa_\lambda=1$ could therefore strongly increase the expected rate and make an observation of the HH process possible earlier than expected in the SM.
The figure also shows the SM differential production cross-section as a function of the HH mass. The strong negative interference is quantified (Fig.~\ref{fig:boxtriangle}~\cite{Collaboration:2759320}).

\begin{figure}[htbp]
\begin{center}
\includegraphics[width=0.41\textwidth]{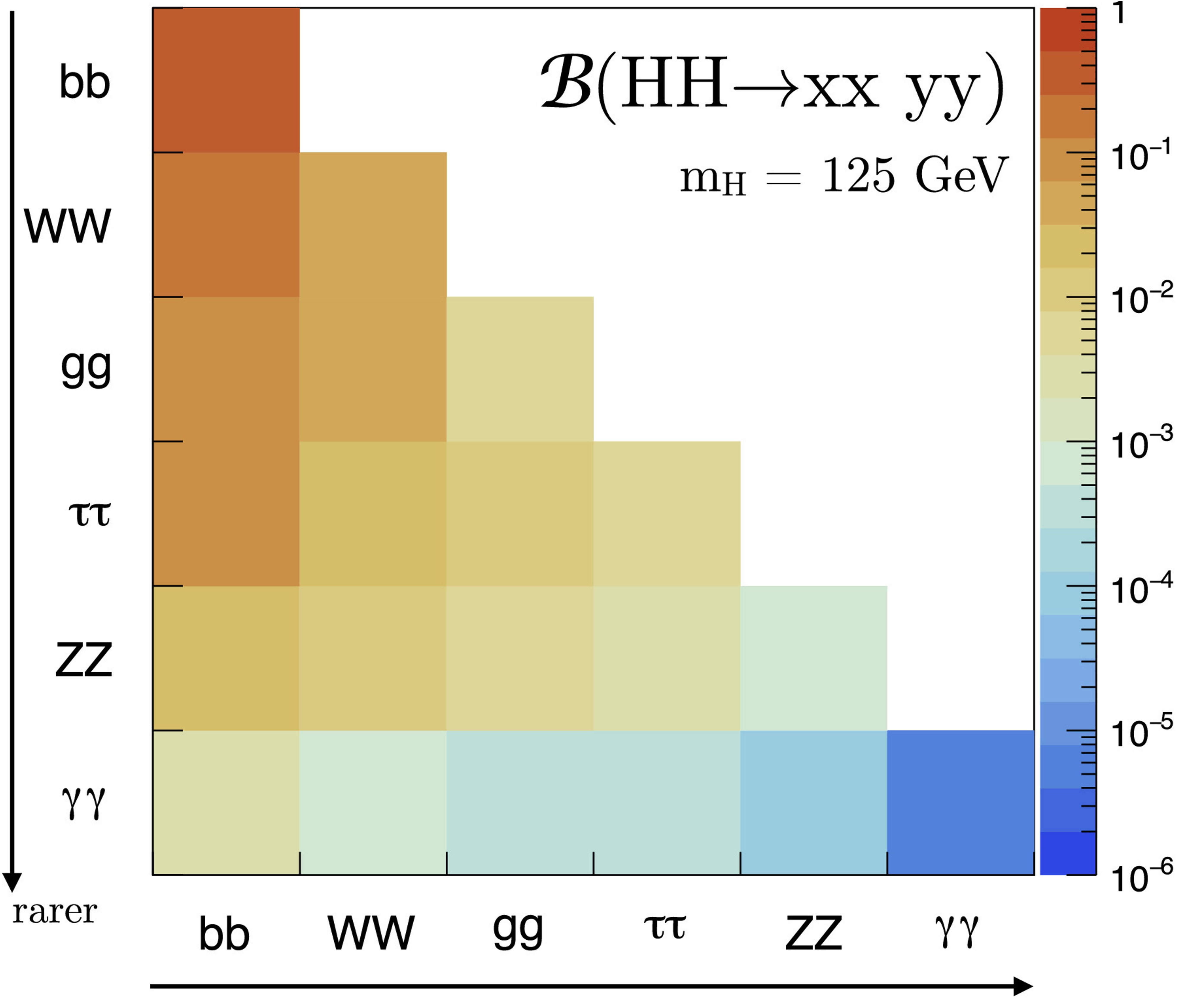}
\includegraphics[width=0.58\textwidth]{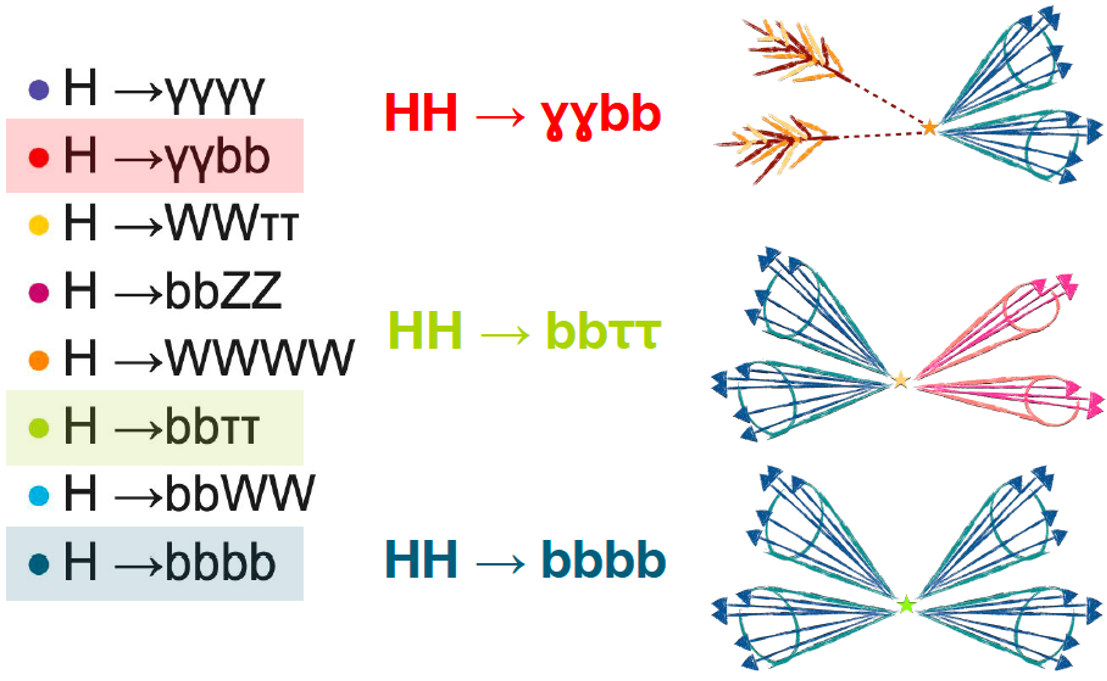}
\end{center}
\vspace{-6mm}
\caption{\label{fig:branching}
Di-Higgs boson decay fractions in the SM, and illustration of expected signatures with b-quark jets, photons, and tau leptons.
}
\vspace{2mm}
\end{figure}

\clearpage
Di-Higgs boson decay fractions in the SM are determined from the SM Higgs boson decay branching fractions. Figure~\ref{fig:branching} quantifies the values.
The figure also illustrates the expected signatures with b-quark jets, photons, and tau leptons.
Owing to the dominant decay branching fraction $H\rightarrow bb$, the HH decay modes with one or both Higgs bosons decaying into a $b$-quark pair lead to the highest HH decay modes.

The previous European Strategy Update (2019) results, summarized in Table~\ref{fig:hh2019}~\cite{Cepeda:2019klc}
concluded:
    “A combined significance of 4 standard deviation can be achieved with all systematic uncertainties included."
This projection on the significance of the HH cross-section with the HL-LHC dataset was based on partial Run-2 analyses (36\,fb$^{-1}$).
This is conservative, due to more recent HH results, thus, there was a need for a new, combined projection.

\begin{table}[htbp]
\vspace{-1mm}
\begin{center}
\includegraphics[width=0.8\textwidth]{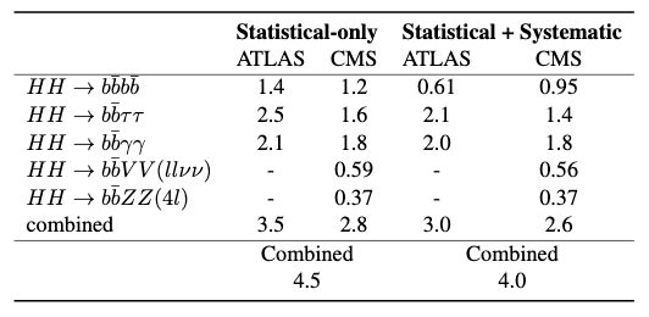}
\end{center}
\vspace{-6mm}
\caption{\label{fig:hh2019}
Previous 2019 discovery prediction for the HH production cross-section and sensitivity contributions from various channels.
From~\protect\cite{Cepeda:2019klc}.
\vspace{5mm}
}
\end{table}
The new results from ATLAS and CMS for the HL-LHC projections~\cite{ATL-PHYS-PUB-2025-018} are based on the full LHC Run-2 analyses (about 5\% of the eventual total LHC dataset), and 
four times more data than were used in 2019 for the previous European Strategy input.
Improved analysis techniques, object, and event reconstruction are taken into account. Furthermore, the combined analysis includes the $\gamma\gamma bb$, $bb\tau\tau$, $bbbb$, $bb\ell\ell$, and multi-lepton channels. 

Various scenarios for systematic uncertainties are compared, in particular:
\begin{itemize}
\item S2: reduced experimental uncertainties, theoretical uncertainties halved, MC statistical uncertainty is neglected.
\item S3: S2 and improved object identification based on Run-3 experience.
\end{itemize}

\begin{figure}[htbp]
\begin{center}
\includegraphics[width=0.43\textwidth]{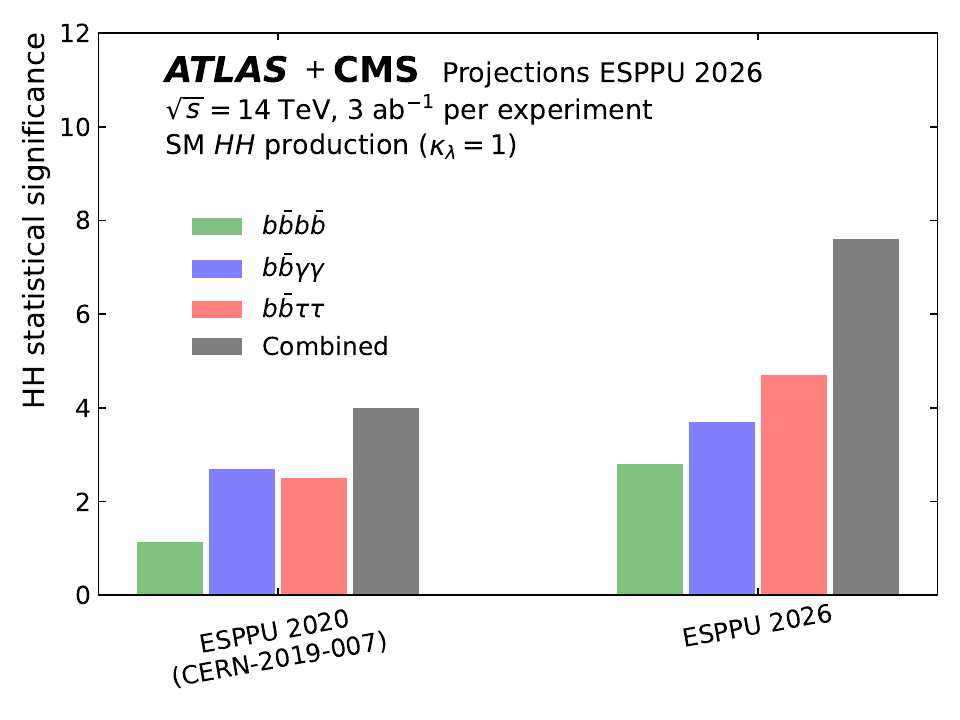}
\includegraphics[width=0.56\textwidth]{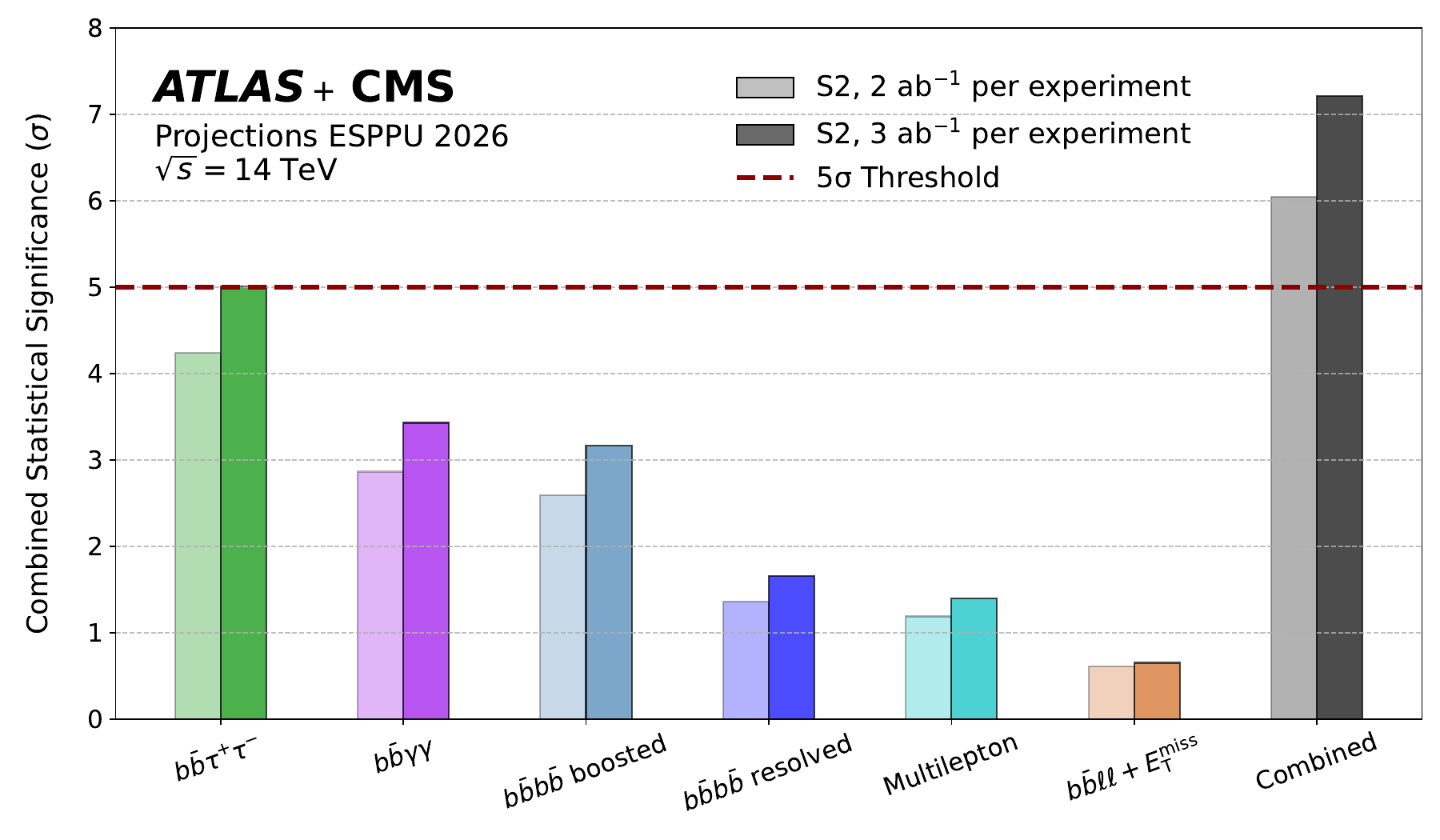}
\end{center}
\vspace{-4mm}
\caption{\label{fig:hh25}
New discovery prediction for the HH production cross-section and sensitivity contributions from various channels.
From~\protect\cite{ATL-PHYS-PUB-2025-018}.
}
\end{figure}

The new ATLAS and CMS projections for the HL-LHC lead to a discovery prediction for the HH production cross-section, as detailed in Fig.~\ref{fig:hh25}~\cite{ATL-PHYS-PUB-2025-018}. The figure also shows the sensitivity contributions from various channels.
The $bb\tau\tau$ channel gives the highest sensitivity contribution. This reminds us of Higgs boson searches at LEP for $hA\rightarrow bb\tau\tau$ having a strong contribution in the BSM Higgs boson searches. 

Table~\ref{fig:hh2025}~\cite{ATL-PHYS-PUB-2025-018} gives details of the contributions of the HH analysis channel of the 
expected statistical significance for HH production, and the
corresponding 68\% confidence interval on $\kappa_3$, assuming $\kappa_3^{\rm true}=1$.
The expected sensitivity is 
\begin{itemize}
    \item S2 2 ab$^{-1}$: 6.0 s.d.,
   \item S2 3 ab$^{-1}$: 7.2 s.d., and
    \item S3 3 ab$^{-1}$: 7.6 s.d.
\end{itemize}
The determination of the Higgs boson self-coupling $\kappa_3$ at 68\% CL is
\begin{itemize}
    \item S2 2 ab$^{-1}$: -32\% / +37\%,
    \item S2 3 ab$^{-1}$: -27\% / +31\%, and
    \item S3 3 ab$^{-1}$: -26\% / +29\%.
\end{itemize}
\begin{table}[htbp]
\begin{center}
\includegraphics[width=\textwidth]{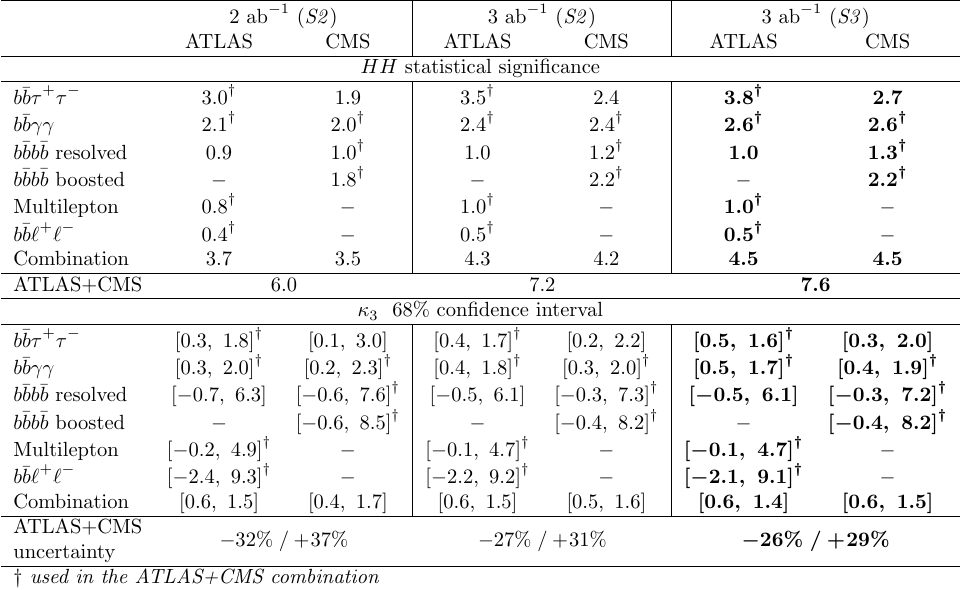}
\end{center}
\vspace{-4mm}
\caption{\label{fig:hh2025}
Discovery prediction for the HH production cross-section and sensitivity contributions from various channels.
From~\protect\cite{ATL-PHYS-PUB-2025-018}.}
\end{table}

\begin{figure}[htbp]
\begin{center}
\includegraphics[width=0.7\textwidth]{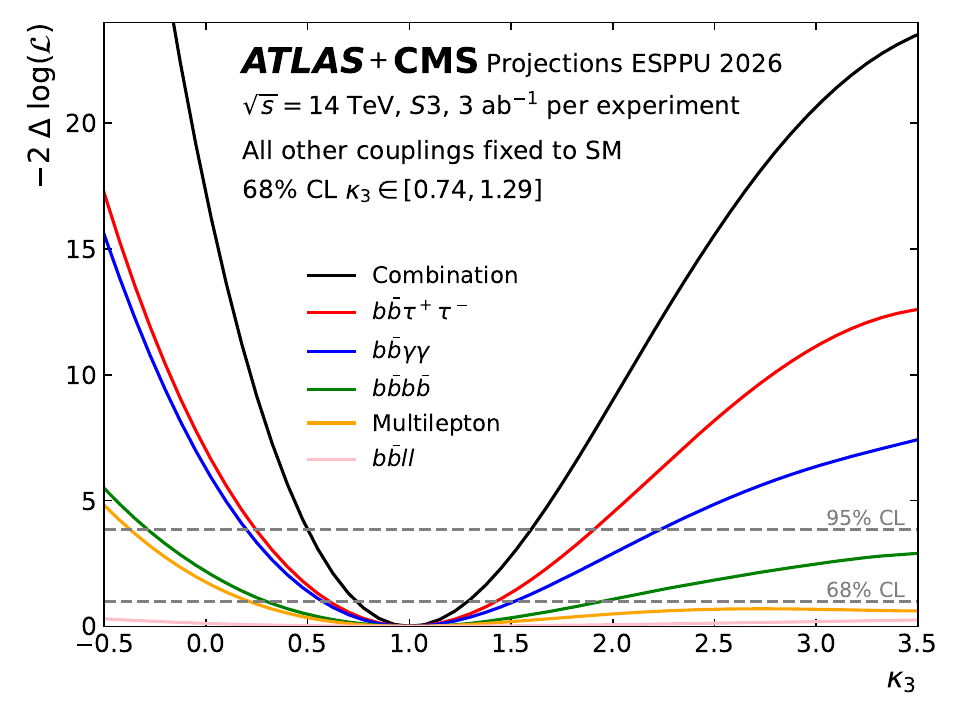}
\end{center}
\vspace{-7mm}
\caption{\label{fig:kappa3}
Contributions to the $\chi^2$ distribution of various analysis channels and their combination as a function of $\kappa_3$ for the S3 scenario 3\,ab$^{-1}$.
From~\protect\cite{ATL-PHYS-PUB-2025-018}.
}
\vspace{-6mm}
\end{figure}

Figure~\ref{fig:kappa3}~\cite{ATL-PHYS-PUB-2025-018} shows the contributions to the $\chi^2$ distribution of various analysis channels and their combination
as a function of $\kappa_3$ for the S3 scenario 3\,ab$^{-1}$.

\clearpage
\subsection{Limitations of HL-LHC projections}
There are several limitations of the HL-LHC projections:
\begin{itemize}
    \item 
Based on Run-2 analyses, it is not possible to incorporate analysis improvements that would be accessible only with higher statistics.
\item Impact of better reconstruction techniques.
\item
Hard to study larger acceptance due to detector upgrades.
\item Impact of trigger upgrades for increased signal acceptance.
\end{itemize}

Therefore, it is difficult to quantify the impact without recorded data, because one can study the increased signal acceptance, but not the extra backgrounds in the high-pileup environment.
Furthermore, additional detector upgrades are also expected to make an impact.

\begin{figure}[htbp]
\begin{center}
\includegraphics[width=0.75\textwidth]{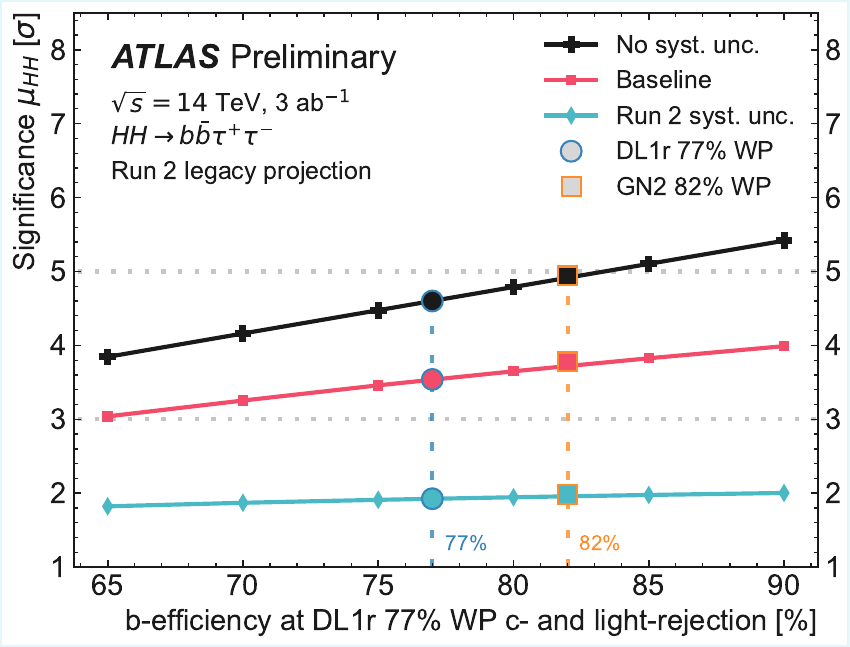}
\vspace{-7mm}
\end{center}
\caption{\label{fig:btag}
Dependence of the HH detection significance as a function of the b-tagging efficiency for the scenarios with Run-2 systematic uncertainties, the baseline (S3) and without systematic uncertainties.
From~\protect\cite{ATLAS:2024voi}.
}
\end{figure}

For example, the $b$-tagging efficiency will have a large impact on the detection significance, as detailed in Fig.~\ref{fig:btag} (from~\cite{ATLAS:2024voi}. For the  $HH\rightarrow bb\tau\tau$ reaction, the figure shows the dependence of the HH detection significance as a function of the $b$-tagging efficiency for the scenarios with Run-2 systematic uncertainties, the baseline (S3) and without systematic uncertainties.

\begin{figure}[htbp]
\begin{center}
\includegraphics[width=0.7\textwidth]{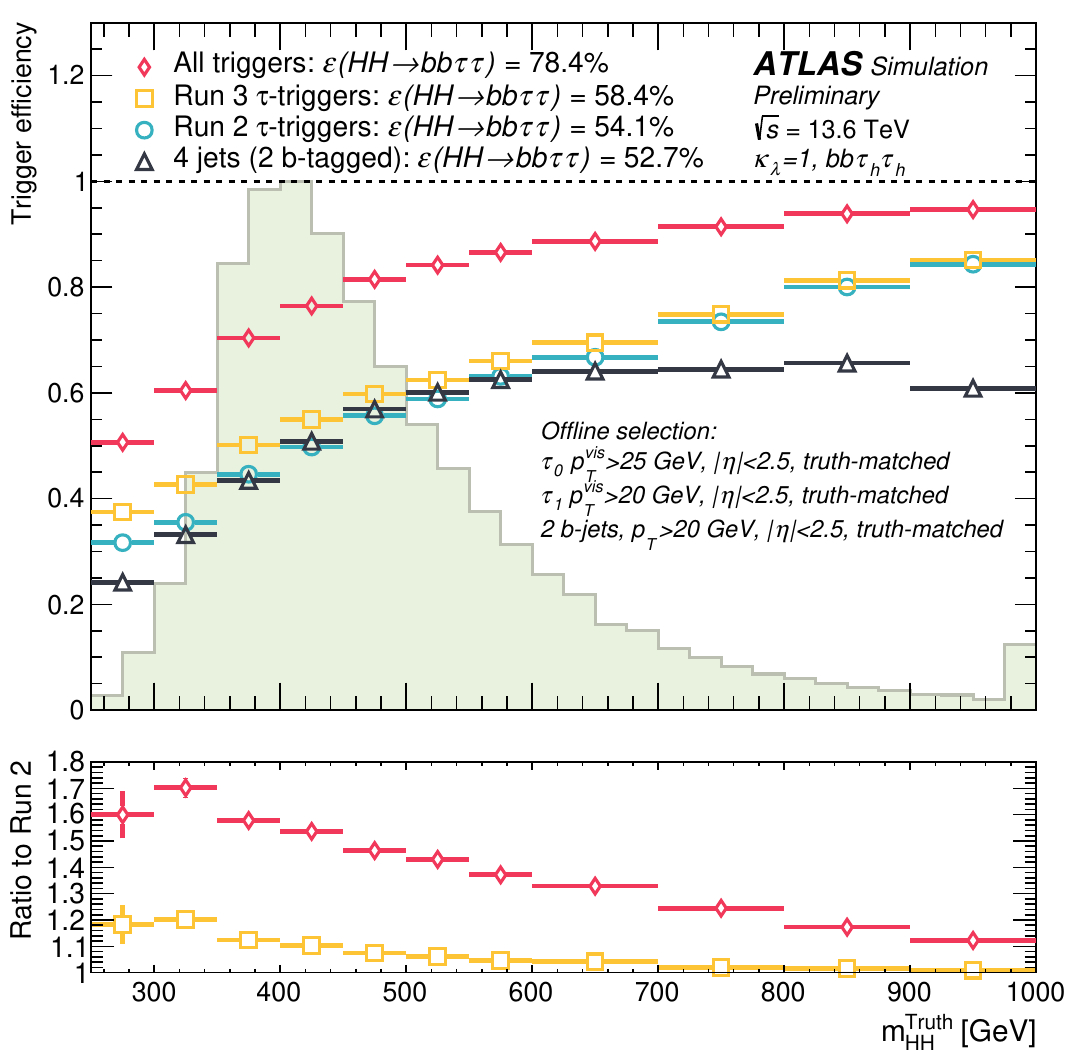}
\vspace{-5mm}
\end{center}
\caption{\label{fig:trigger}
Trigger efficiencies in a simulated event sample of SM Higgs boson pair production at a centre-of-mass energy of 13.6\,TeV, with decays to two $b$-quarks and two hadronically decaying $\tau\tau$-leptons as a function of the generated Higgs pair mass ($m_{HH}^{\rm truth}$). 
From~\protect\cite{ATL-trigger}.
}
\vspace{-4mm}
\end{figure}

Also, the tau trigger performance will have a significant impact on the trigger efficiency for $bb\tau\tau$ analysis channels, as shown in Fig.~\ref{fig:trigger}~\cite{ATL-trigger}.
The figure shows the trigger efficiency as a function of the generated Higgs pair mass ($m_{HH}^{\rm truth}$), increasing from about 50\% to 95\% for all triggers.

\section{Beyond Standard Model (BSM)}

In models beyond the SM, the trilinear coupling (self-coupling) could have significantly different values from the SM model unity.  Thus, the linearity of the self-coupling measurement with respect to the true value is addressed. Also, at the HL-LHC, the shape of the Higgs potential and First-Order Phase Transition (FOPT) can be constrained.
The extension of the SM with extra scalars has been studied.
This enriched scalar potential dynamics enables the possibility of a strong FOPT as it interplays with the Higgs potential.
Furthermore, future precision mass measurements will give insight into the vacuum stability in the Universe.

\subsection{Linearity of the self-coupling measurement}

\begin{figure}[htbp]
\vspace{-2mm}
\begin{center}
\includegraphics[width=0.46\textwidth]{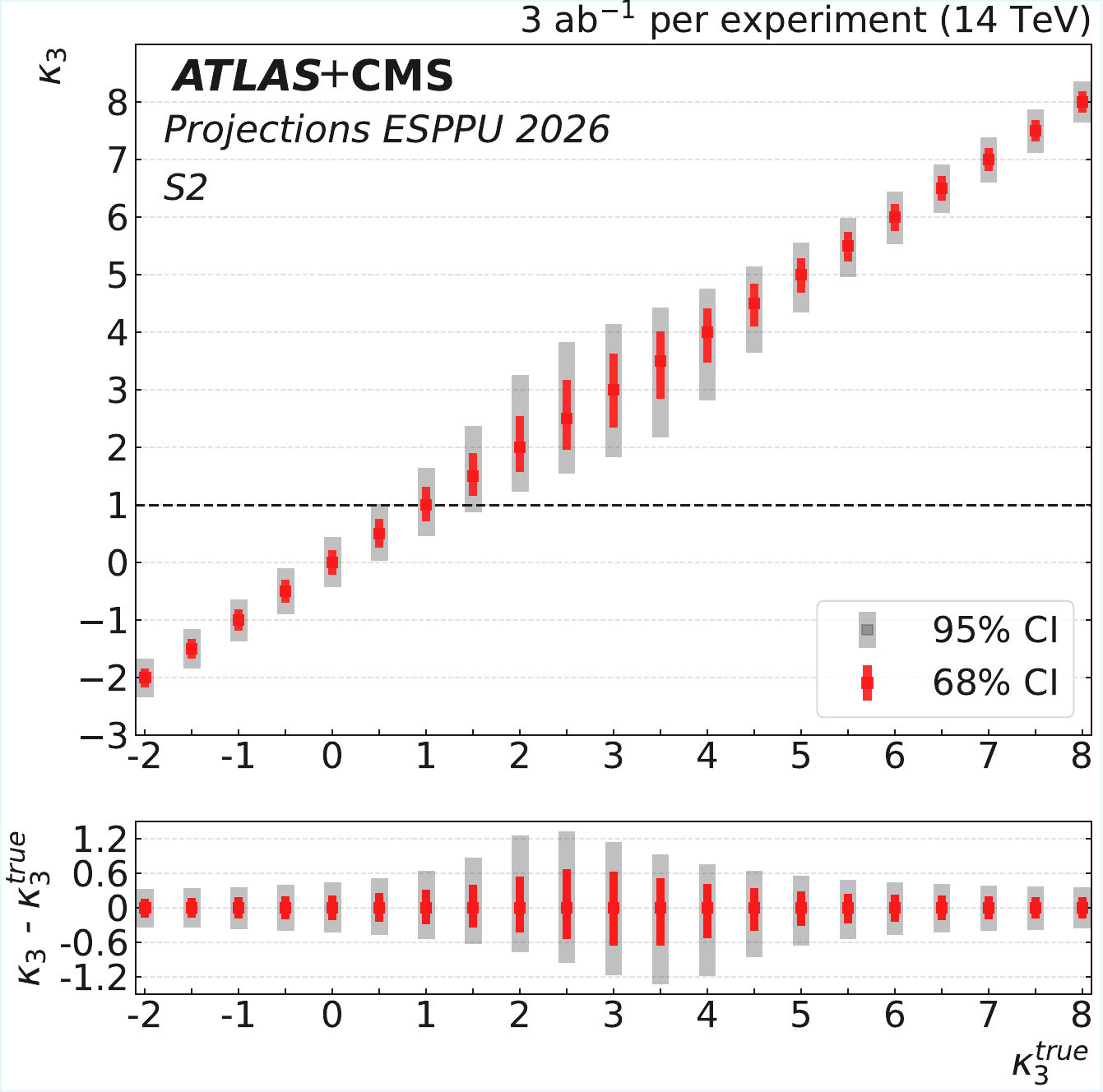}
\includegraphics[width=0.53\textwidth]{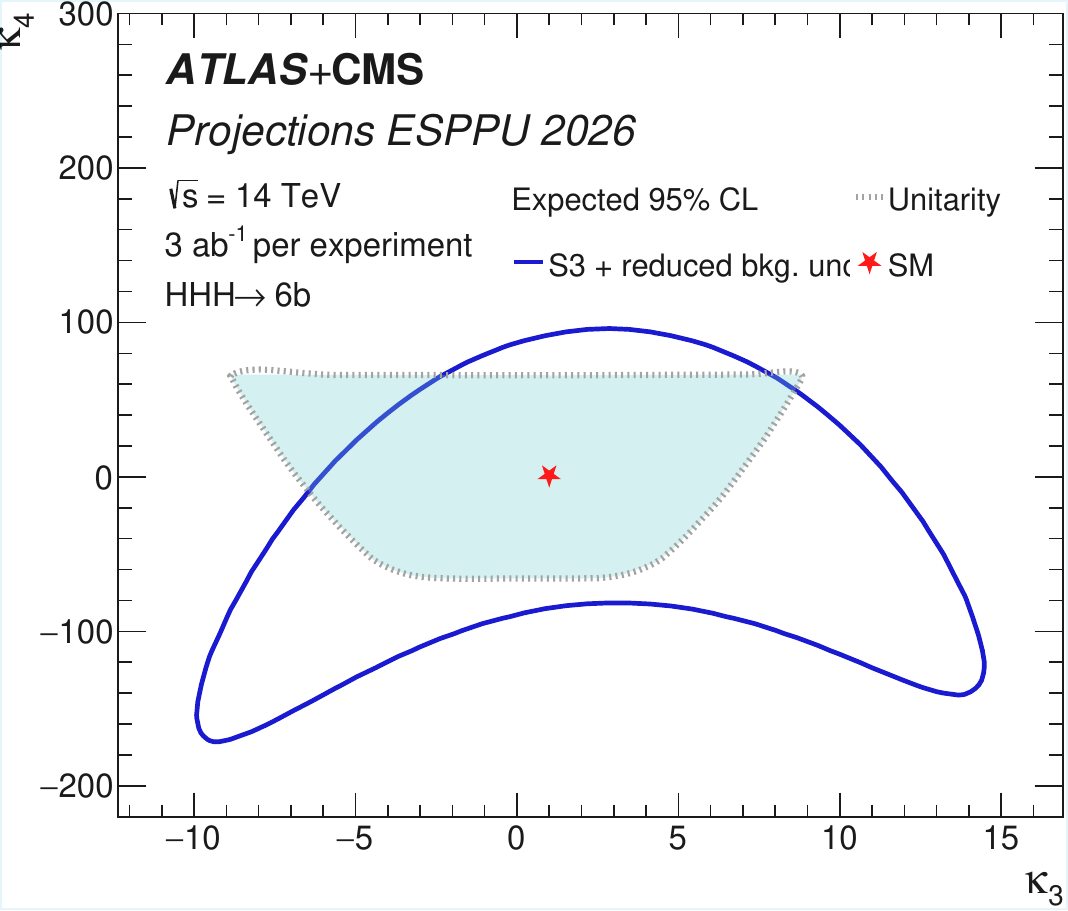}
\vspace{-10mm}
\end{center}
\caption{\label{fig:linear}
Left: Expected $\kappa_3$ measurement and uncertainty as a function of the $\kappa_3^{\rm true}$ values.
Right: Constraints from the HHH search projection on $\kappa_3$ and $\kappa_4$.
From~\protect\cite{ATL-PHYS-PUB-2025-018}.
}
\vspace{-10mm}
\end{figure}

For the future measurement of $\kappa_3$, it is important to study the linearity between the true $\kappa_3$ value and the expected measurement. Figure~\ref{fig:linear}~\cite{ATL-PHYS-PUB-2025-018} shows the linearity and the expected measurement uncertainties. 
The projected combined results from ATLAS and CMS are presented assuming 3\,ab$^{-1}$ per experiment in the S2 scenario.
Good linearity between the true and measured $\kappa_3$ values within the measurement uncertainty is expected.
Figure~\ref{fig:linear}~\cite{ATL-PHYS-PUB-2025-018}
also shows constraints from an HHH search projection on $\kappa_4$. In the S3 scenario with 3\,ab$^{-1}$, the expected limit at 95\% CL is 86 times the SM expectation~\cite{ATL-PHYS-PUB-2025-018}.

\subsection{BSM Higgs boson potential measurement}

\begin{figure}[htbp]
\begin{center}
\includegraphics[width=0.9\textwidth]{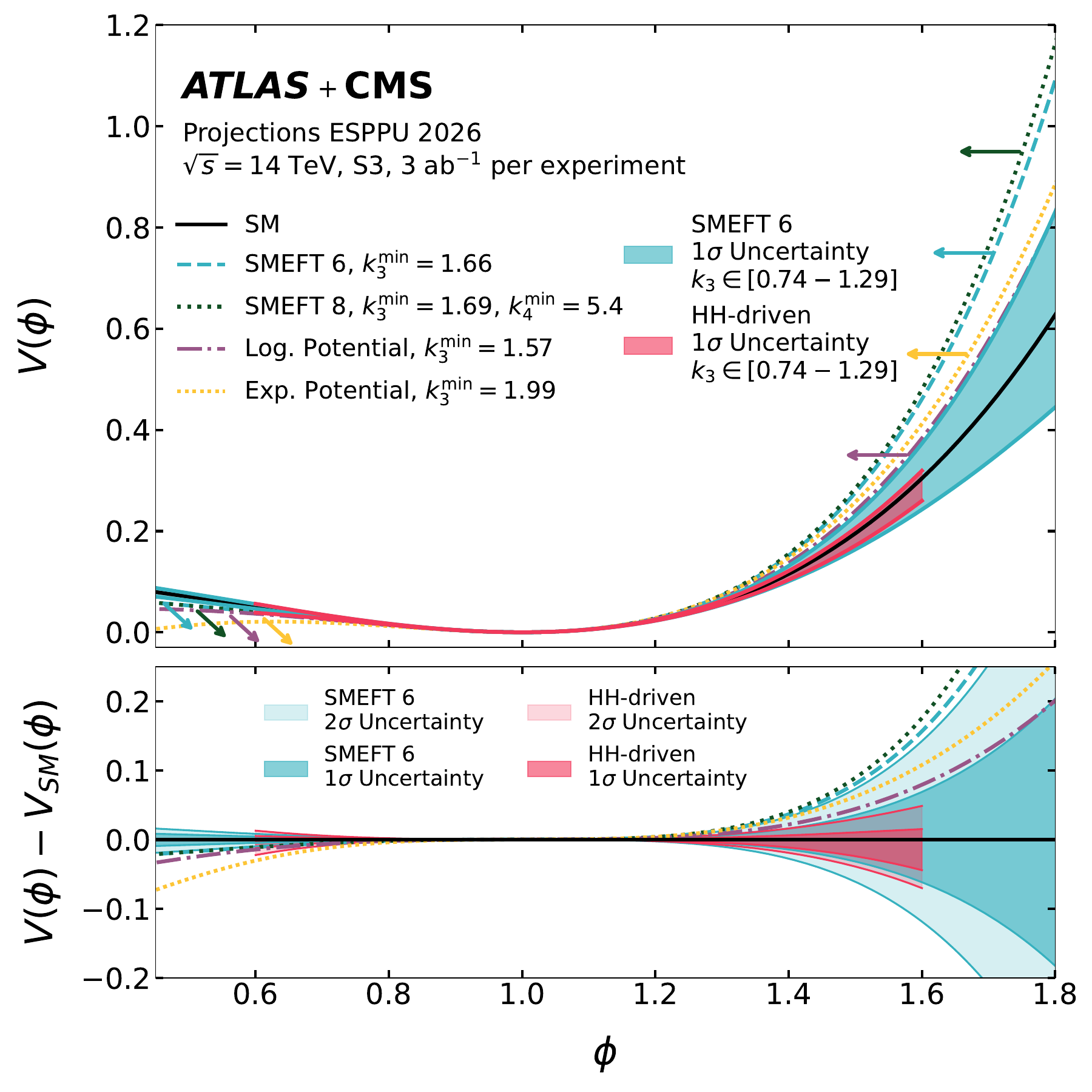}
\vspace{-7mm}
\end{center}
\caption{\label{fig:potential2}
Higgs potentials in various models which predict first-order phase transition. The models are compared with the SM Higgs potential. Two approaches (SMEFT 6 and HH-driven) are used to show the expected uncertainties on the Higgs self-coupling achieved by combining ATLAS and CMS at 3\,ab$^{-1}$ in the S2 scenario, in a wide range of the Higgs field value. The bottom panel shows the difference between the potential $V(\Phi)$ and its SM expectation $V_{\rm SM}(\Phi)$. Here, the 68\% and 95\% CL uncertainty bands of the shape of $V(\Phi)$ are shown, for the HH-driven and SMEFT 6 potentials. The bottom plot shows the zoom into the $V_{\rm SM}(\Phi)$ difference around the minimum of $V(\Phi)$, corresponding to the validity range of the HH-driven band. From~\protect\cite{ATL-PHYS-PUB-2025-018}.
}
\end{figure}

A central question is whether an SM Higgs potential or an alternative potential is realized in Nature. 
For the S2 scenario with 3\,ab$^{-1}$, the SM potential and the expected uncertainty of its measurement are shown in Fig.~\ref{fig:potential2}~\cite{ATL-PHYS-PUB-2025-018}.
The four non-solid lines correspond to the
four alternative potentials and display the limit of the variation of the shapes that could still imply a strong FOPT in the early Universe. These lines are obtained for values of the parameters as indicated in the legend of the figure.
Only variations of the shapes in the direction indicated by the arrows on the four curves would allow for a strong
FOPT. In other words, the arrows on each of these curves indicate the direction in which the modified potentials
implying a strong FOPT lie on the plot.
The darker blue and red shaded areas show the allowed band for the Higgs potential at 68\% CL derived
from the ATLAS and CMS projections.
The red area illustrates the allowed range on the third derivative of the
potential (which is sensitive to $\lambda_3$), ignoring higher-order terms.
Its finite range around the minimum aims at representing the fact that LHC constraints coming from
HH production can only determine $\kappa_3$ (at LO) in a model-independent way close to the minimum. It is therefore labelled “HH-driven”, as it represents what the experiments are directly sensitive to.
The light blue and light red contours represent the SM potential variations corresponding to the 68\% and 95\% CL uncertainty bands on $\kappa_3$.

\subsection{Precision Higgs physics and searches for a heavy scalar singlet}

\begin{figure}[htbp]
\begin{center}
\includegraphics[width=0.49\textwidth]{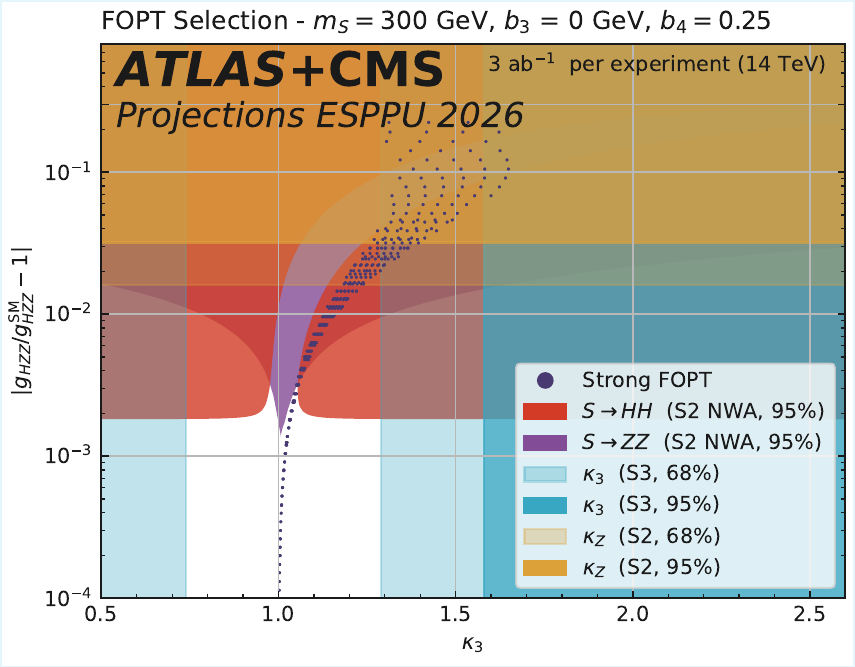}
\includegraphics[width=0.49\textwidth]{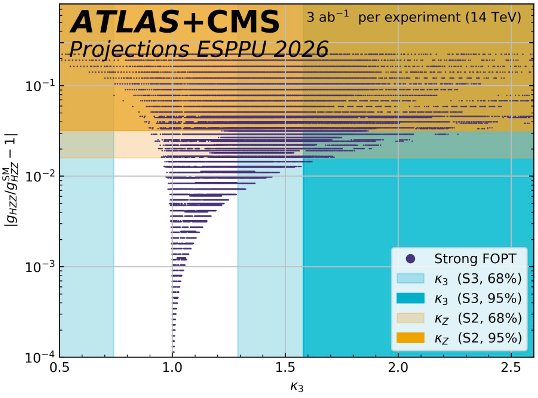}
\vspace{-7mm}
\end{center}
\caption{\label{fig:singlet}
Left: Exclusion bounds in the plane of the deviation of the Higgs boson coupling to the $Z$ with respect to the SM one as a function of $\kappa_3$.
The dark blue points show the area where a strong FOPT in the early Universe is possible within the scalar singlet model for $m_S = 300$\,GeV.
Right:
Exclusion bounds in the plane of the Higgs boson to $ZZ$ coupling with respect to the SM one as a function of $\kappa_3$. 68\% and 95\% exclusion bounds are displayed. The dark blue points populate the area where a strong FOPT in the early Universe is possible within the scalar singlet model.
From~\protect\cite{ATL-PHYS-PUB-2025-018}.
}
\end{figure}

Precision Higgs physics interplays with searches for a heavy scalar singlet.
Some minimal extensions of the SM involve the inclusion of a new real scalar singlet, $S$, which couples to the SM via the Higgs portal. The extended 
Higgs boson potential is

\begin{equation}
\begin{split}
V(h) 
& =\frac{1}{2} m_H^2h^2 
+ \lambda v h^3 
+ \frac{1}{4}\lambda h^4 
+ b_1 S\\
&+ \frac{1}{2}\mu_S^2
+ \frac{1}{4}b_4S^4
+ \frac{1}{2}a_2|\Phi|^2S^2
+ \frac{1}{3} b_3 S^2
+ \frac{1}{2}a_1|\Phi|^2S.
\end{split}
\end{equation}
If the additional scalar is sufficiently heavy, it can decay into a pair of Higgs bosons.
This could enable the possibility of a strong FOPT.
The stability of the vacuum for large field configurations can be affected~\cite{ATL-PHYS-PUB-2025-018}.
Figure~\ref{fig:singlet}~\cite{ATL-PHYS-PUB-2025-018} shows
exclusion bounds in the plane of the deviation of the Higgs boson coupling to the $Z$ with respect to the SM one as a function of $\kappa_3$.
Dark blue points show the area where a strong FOPT in the early Universe is possible within the scalar singlet model for $m_S=300$\,GeV.
Scenarios S2 and S3 for 3\,ab$^{-1}$ data per experiment and combined ATLAS and CMS results are shown.
The figure also shows 
exclusion bounds in the plane of the Higgs boson to $ZZ$ coupling with respect to the SM one as a function of $\kappa_3$, and the coverage of 
points where a
FOPT in
the early Universe
is possible within
the scalar singlet
model~\cite{ATL-PHYS-PUB-2025-018}.
Scenarios S2 and S3 are presented for 
3\,ab$^{-1}$ data per experiment and combined ATLAS and CMS results.

\subsection{Projections of the Higgs boson and top quark masses}

\begin{figure}[htbp]
\begin{center}
\includegraphics[width=0.8\textwidth]{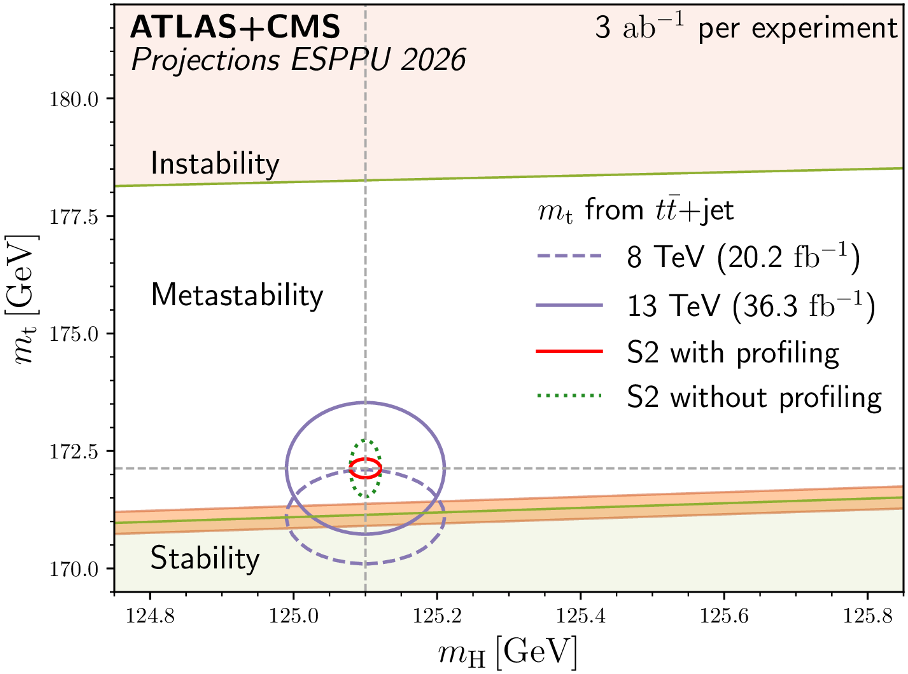}
\vspace{-7mm}
\end{center}
\caption{\label{fig:mtop}
Contours in the plane $m_t$ versus $m_H$. Instable, metastable, and stable vacuum regions are shown. 
The band between the stable and metastable
region represents the uncertainty in $\alpha_S$. From~\protect\cite{ATL-PHYS-PUB-2025-018}.
}
\end{figure}

Projections for the HL-LHC measurements of the Higgs boson and top quark masses relate to the stability of the vacuum in the Universe.
The $m_H$ measurement projections are based on 3\,ab$^{-1}$ of integrated luminosity per experiment in the S2 scenario (Fig.~\ref{fig:mtop}~\cite{ATL-PHYS-PUB-2025-018}).
These projections result in a total uncertainty on $m_H$ 
from $H\rightarrow ZZ^* \rightarrow 4\ell~(\ell = e~{\rm or}~\mu$) events of 38\,MeV for ATLAS and 26\,MeV for CMS, leading to a combined uncertainty of 21\,MeV~\cite{ATL-PHYS-PUB-2025-018}.

\section{Past-present-future LHC operation}

\begin{figure}[tp]
\vspace{-6mm}
\begin{center}
\includegraphics[width=0.77\textwidth]{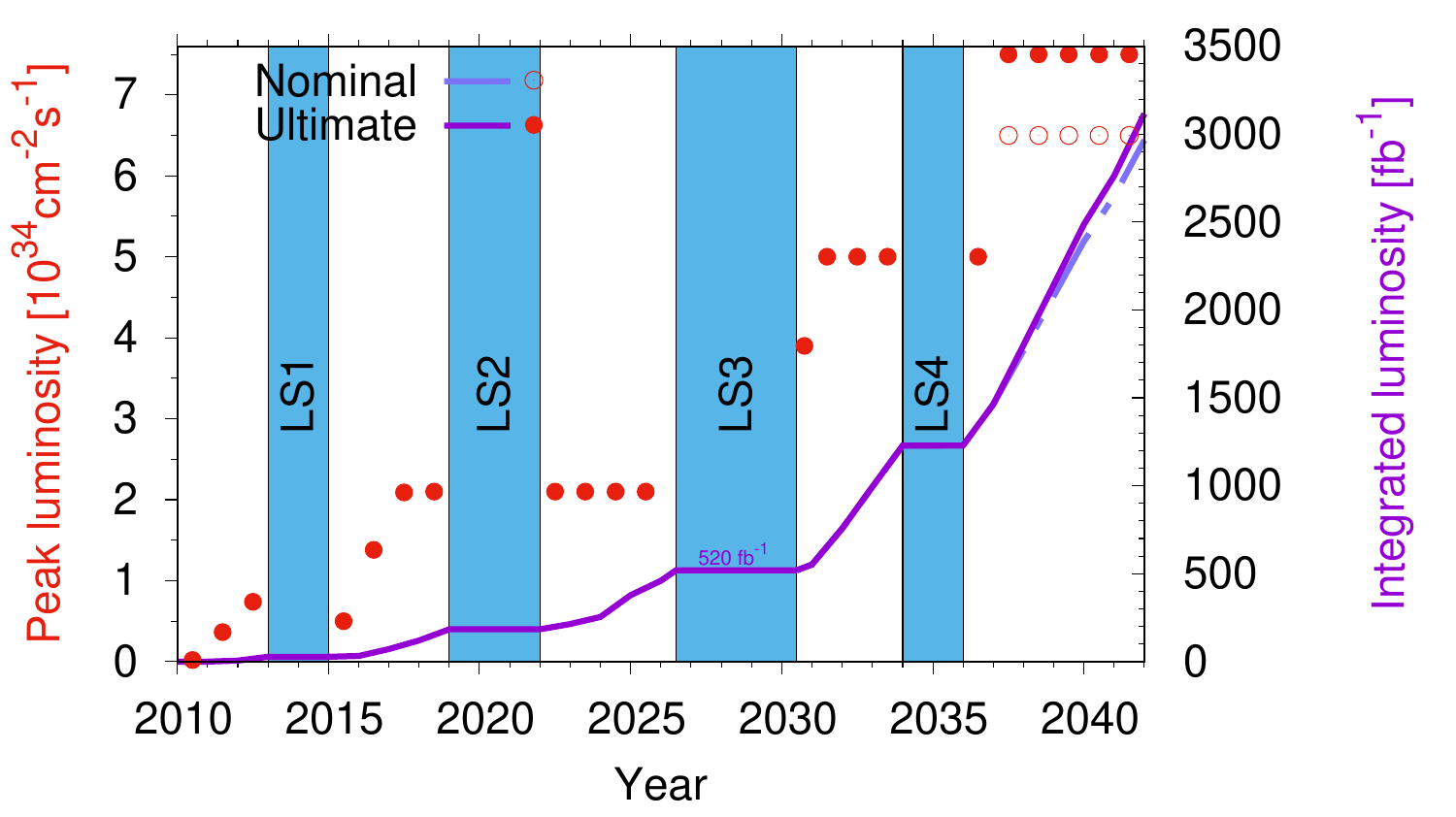}
\vspace{-7mm}
\end{center}
\caption{\label{fig:future}
Overview and details of the past, present, and future LHC operations. From~\protect\cite{ATL-UPGRADE-PUB-2025-001}.
}
\vspace{-6mm}
\end{figure}

Figure~\ref{fig:future}~\cite{ATL-UPGRADE-PUB-2025-001} gives an overview  of the past, present, and future LHC operation schedule.
Currently, LHC-3 is taking data. In 2026, the Long Shutdown 3 (LS3) will start. During LS3, installations for the HL-LHC operation will take place.
Data taking will start after a commissioning phase in 2030, then protons will be colliding for high-luminosity physics analyses.
The LHC operation for Run-4 is scheduled for a duration of four years. Afterwards, the Long Shutdown 4 (LS4) is planned before Run-5 starts. The HL-LHC planning extends beyond 2040.

\clearpage

\section{Conclusions}

LHC Higgs predictions for the European Strategy for Particle Physics are updated for the 2026 report.
The $H \rightarrow \mu\mu$, precision of 3\% and
$H\rightarrow Z\gamma$, and precision 7\% are expected.
Other Higgs boson couplings precisions range between  1.6\% to 3.6\%.
The differential cross-sections can be measured with good sensitivity in the 
high-$p_T$(Higgs) region.
Projections for HH measurements in various final states and their combination are updated.
An HH significance of more than $7\sigma$ is expected for the HL-LHC.
Better than 30\% precision on SM self-coupling is expected, based on direct HH \mbox{ATLAS} and CMS Run-2 results, and nearly a single-experiment observation can be achieved.
However, the reach of the HHH production search is 86 times the SM expectation at 95\% CL.
Regarding constraining the shape of the electroweak vacuum potential, the vast majority of the parameter space will be covered.
Precision measurements of the top quark mass will reach an accuracy of 200\,MeV, combined with the Higgs boson mass precision of 21\,MeV, enabling exploration of the nature of the electroweak vacuum and assessment of the stability of the Universe.
These projections are still conservative, as upcoming detector and trigger upgrades, along with advances in machine learning analysis techniques, are expected to further enhance performance.
The updated physics goals of the HL-LHC are in line with the phenomenological studies.
Several results are limited by theoretical uncertainties, highlighting the need for further progress in high-precision theoretical calculations aligned with the demands of the HL-LHC.

\vspace{-3mm}
\section*{Acknowledgements}
The research is supported by the Ministry of
Education, Youth and Sports of the Czech
Republic under the project number LM 2023040.
Copyright 2025 CERN for the benefit of the ATLAS Collaboration. CC-BY-4.0 license.

\clearpage
\bibliographystyle{unsrt}
\bibliography{lib}

\end{document}